# Algorithmic Superactivation of Asymptotic Quantum Capacity of Zero-Capacity Quantum Channels

Laszlo Gyongyosi[*], Sandor Imre

*Department of Telecommunications*

*Budapest University of Technology and Economics*

H-1111 Budapest, 2 Magyar tudosok krt, Hungary

[*]*gyongyosi@hit.bme.hu*

**The superactivation of zero-capacity quantum channels makes it possible to use two zero-capacity quantum channels with a positive joint capacity for their output. Currently, we have no theoretical background to describe all possible combinations of superactive zero-capacity channels; hence, there may be many other possible combinations. In practice, to discover such superactive zero-capacity channel-pairs, we must analyze an extremely large set of possible quantum states, channel models, and channel probabilities. There is still no extremely efficient algorithmic tool for this purpose. This paper shows an efficient algorithmical method of finding such combinations. Our method can be a very valuable tool for improving the results of fault-tolerant quantum computation and possible communication techniques over very noisy quantum channels.**

## 1. Introduction

This paper introduces an extremely efficient algorithmic solution for finding superactive zero-capacity channels and provides an algorithmic framework for analyzing the properties of channel output states in quantum space. In 2008, Smith and Yard have found only one possible combination for superactivation [52], and there should be many other possible combinations. Here, we confirm this with our method, which can be extended to discover other such combinations of channels. The number of efficient approximation algorithms for quantum informational distances is very small because of the special properties of quantum informational generator functions and of asymmetric quantum informational distances. If we wish to analyze the properties of quantum channels using today's classical computer architectures, an extremely efficient algorithm is needed. The proposed method gives an algorithmic solution to the superactivation problem of zero-capacity channels. We exhibit a fundamentally new and efficient algorithmic solution for discovering all possible "superactive" channel combinations. As we will show, Smith's result [52] is only one possible solution to superactivation, and many other solutions can be discovered by our algorithmic solution.

This paper is organized as follows. In Section 2, we present the theoretical background of the proposed approach. In Sections 3 and 4, we analyze the geometry of quantum states and the problem of superactivation. In Section 5, we give a fundamentally new geometrical interpretation of the superactivation of asymptotic quantum capacity. In Section 6, we give an illustrative example. Finally, in Section 7, we present our conclusions.

**1.1 Related Work**

The superactivation of the quantum capacity of zero-capacity quantum channels was introduced by Smith and Yard [52]. Since the revolutionary properties of superactivation quantum channel capacities were first reported on, many further quantum informational results have been achieved [9,14,53], [58-61]. Recently, Duan and Cubitt et al. found a possible combination for the superactivation of the classical zero-error capacity of quantum channels, which has opened up a debate regarding the existence of other possible channel combinations [9,14]. The problem of superactivation can be discussed as part of a larger problem – the problem of quantum channel additivity. The additivity problem is considered to be a very important problem in quantum information theory.

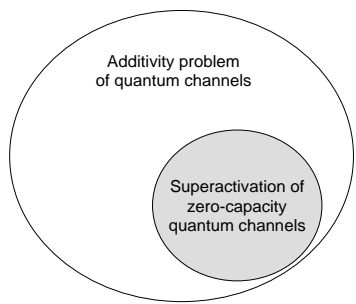

**Fig. 1.** The problem of superactivation of zero-capacity quantum channels as a sub domain of a larger problem set.

The interpretation of quantum channel capacity based on the quantum relative entropy function was theoretically presented by Cortese [8], Hayashi et al. [32,33], and Schumacher-Westmoreland [48,49]. However, the most basic questions regarding the different types of capacity of a quantum channel remain open. Most types of quantum channel capacity have been found to be non-additive [5,12,16,18,50,51,55].

In this paper, we exhibit a fundamentally new geometrical approach to finding *superactive* zero-capacity quantum channels. For the combination of any quantum channel $\mathcal{N}_1$ that has some private capacity $P(\mathcal{N}_1) > 0$ and of a "fixed" 50% erasure symmetric channel $\mathcal{N}_2$, the following connection holds between the asymptotic quantum capacity $Q(\mathcal{N}_1 \otimes \mathcal{N}_2)$ of the joint structure $\mathcal{N}_1 \otimes \mathcal{N}_2$, and the private capacity $P(\mathcal{N}_1)$ of $\mathcal{N}_1$:

$$Q(\mathcal{N}_1 \otimes \mathcal{N}_2) \geq \frac{1}{2} P(\mathcal{N}_1). \tag{1}$$

The channel combination for the superactivation of the asymptotic quantum capacity of zero-capacity quantum channels is shown in Fig. 2.

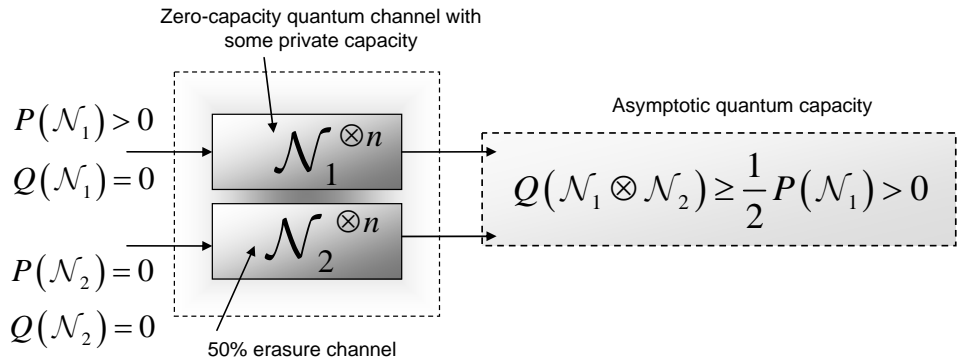

**Fig. 2.** The first channel has some positive private capacity, and the second quantum channel is a 50% erasure channel with zero quantum capacity.

In what follows, we will see that it is possible to find other combinations of quantum channels $\mathcal{N}_1$ and $\mathcal{N}_2$, which are, individually, "zero-capacity" in the sense that

$$Q(\mathcal{N}_1) = Q(\mathcal{N}_2) = 0, \tag{2}$$

and yet satisfy

$$Q(\mathcal{N}_1 \otimes \mathcal{N}_2) > 0. \tag{3}$$

We use computational geometric methods since these algorithmic tools can be implemented very efficiently [2,13,15,41,43]. The problem of clustering in quantum space, using the quantum informational distance as a distance function, is a completely new area in quantum information theory. Recently, the possibilities of the application of computational geometric methods in quantum space have been studied by Kato et al. [38] and Nielsen et al. [44] and Nock and Nielsen [45]; however, the problem of clustering was not analyzed in their work. The coreset method for different distances has been studied in the literature. Euclidean methods were studied in [1,6,7,13,31,44] and a non-Euclidean metric by Banerjee [3] and Ackermann et al. in [1]. A very useful and practical approach to the computation of quantum channel capacity using geometric methods was also presented by Hayashi et al. [32,33]. In our work, we will construct advanced algorithmic approaches to analyze the superactivation of the asymptotic quantum capacity of quantum channels [29].

## 2. Computational Geometry in Quantum Information Processing

In our work, we apply computational geometry in quantum space. With the help of efficient computational geometric methods, the superactivation of zero-capacity quantum channels can be analyzed very efficiently. We would like to analyze the properties of the quantum channel using classical computer architectures and algorithms [4,10,17,42,56] since, currently, *we have no quantum computers*. To this day, the most efficient classical algorithms for this purpose are computational geometric methods. We use these classical computational geometric tools to discover the still unknown "*superactive*" zero-error capacity quantum channels.

Computational geometry was originally focused on the construction of efficient algorithms and provides a very valuable and efficient tool for computing hard tasks. In many cases, traditional linear programming methods are not very efficient.

To analyze a quantum channel for a large number of input quantum states with classical computer architectures, very fast and efficient algorithms are required. We will use quantum information as a distance measure instead of classical geometric distances. Unlike ordinary geometric distances, the quantum informational distance is not a metric and is not symmetric. We combine the models of information geometry and the methods of computational geometry. At present, computational geometry algorithms are an active, widely used, and integrated research field [11,37,40,47,46]. Many difficult problems can be solved by computational geometric methods if provided with well-designed and efficient algorithms [1,3,20,21,22,23,24,25,26,27,28,29].

### 2.1. Analysis of Superactivation

To study the geometry of superactivation, we define a new abstract geometric object, the quantum informational *superball*. Using the coreset method [1,15] we can analyze the capacities of quantum channels for an *extremely large* input set and for all possible channel models with extremely high efficiency. These informational geometric algorithms can be applied to quantum informational distances. Then the additivity properties of different quantum channel models with various channel probabilities can be analyzed extremely quickly in our fundamentally new framework. The iterations are made on the channel input states, channel models, and channel parameters. The output of the algorithm is the radius of the quantum informational "superball," which describes the asymptotic quantum capacity of the channel.

## 3. Communication over Quantum Channels

In the classical communication model, the sender and receiver can be modeled by random variables $X = \{p_i = P(x_i)\}$, $i = 1,...N$, and $Y = \{p_i = P(y_i)\}$, $i = 1,...N$. In classical systems, the Shannon entropy of the discrete random variable $X$ is defined as $H(X) = -\sum_{i=1}^{N} p_i \log(p_i)$, where the logarithm function, $\log(\cdot)$ is, as usual in this context, to base 2. The probability of a random variable $X$ given (or, as one says, "conditioned on") $Y$ is denoted by $p(X|Y)$. The noise in the channel increases the uncertainty in $X$, given Bob's output $Y$. The informational theoretic noise of the channel increases the conditional Shannon entropy, defined as $H(X|Y) = \sum_{i=1}^{N_X} \sum_{j=1}^{N_Y} p(x_i, y_j) \log p(x_i|y_j)$; thus, the radius of the smallest enclosing quantum informational ball will decrease for fixed $H(X)$ [35,55].

The general classical informational theoretic model for a noisy quantum channel is as follows. A quantum state can be described by its *density matrix* $\rho \in \mathbb{C}^{d \times d}$, which is a $d \times d$ matrix where $d$ is the level of the given quantum system. For an $n$ qubit system, the level of the quantum system is $d = 2^n$. We use the fact that particle state distributions can be analyzed probabilistically by means of density matrices. A *two-level* quantum system can be defined by its density matrices as follows:

$$\rho = \frac{1}{2}\begin{pmatrix} 1+z & x-iy \\ x+iy & 1-z \end{pmatrix}, \quad x^2 + y^2 + z^2 \leq 1, \tag{4}$$

where $i$ denotes a square root of –1. The density matrix $\rho = \rho(x, y, z)$ can be identified with a *point* $(x, y, z)$ in three-dimensional space, and the ball **B** formed by such points, $\mathbf{B} = \{(x, y, z) | x^2 + y^2 + z^2 \leq 1\}$, is called the Bloch ball. A quantum state $\rho$ can also be represented in spherical coordinates $\rho = (r, \theta, \varphi)$, where $r$ is the distance from the quantum state to the origin and $\theta$ and $\varphi$ represents the latitude and longitude rotation angles, respectively.

A quantum channel $\mathcal{N}$ can be described by an affine map, which maps quantum states to other quantum states. The presence of noise in a quantum channel means that the mapping is no longer a noiseless one-to-one relationship. The image of the quantum channel's transform $\mathcal{N}$ is an ellipsoid [5,8,48]. Geometrically, $\mathcal{N}$ maps the Bloch ball to a deformed ball contained inside the Bloch ball [38]. To preserve the property of being a density matrix $\rho$, the map that models the quantum channel $\mathcal{N}$ must be trace-preserving, i.e., $Tr\mathcal{N}(\rho) = Tr(\rho)$, where $Tr(\cdot)$ is the trace operation, and it must be completely positive; i.e., for the identity map $I$, $\mathcal{N} \otimes I$ maps a semi-positive Hermitian matrix to a semi-positive Hermitian matrix [12,39,51].

In Fig. 3, Alice's pure state is denoted by $\rho_A$, and Bob's mixed input state is denoted by $\mathcal{N}(\rho_A) = \sigma_B$. For random variables $X$ and $Y$, as above, $H(X, Y) = H(X) + H(Y|X)$. We will give a geometric representation of the information that can be transmitted in the presence of noise in a quantum channel.

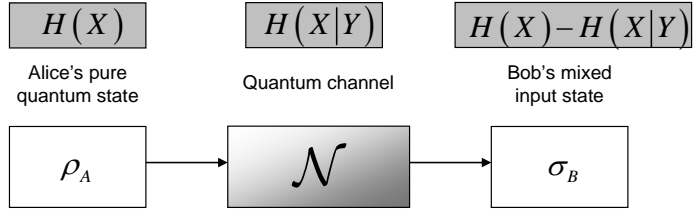

**Fig. 3.** The classical communication model.

In classical information theory, information that can be transmitted through a noisy channel can be measured in a geometric way by the radius of the smallest informational ball that encloses channel output states. We seek to maximize $H(X)$ and minimize $H(X|Y)$ to maximize the radius of the smallest enclosing ball since the radius $r_{classical}$ of the classical informational ball can be computed as:

$$r_{classical} = \max_{\{all\ possible\ x_i\}} H(X) - H(X|Y). \tag{5}$$

In quantum information theory, instead of a classical ball, we have a quantum informational ball that contains all channel output quantum states. The entropies between mixed quantum states are measured by the *von Neumann entropy* [19,36,57] – instead of the classical Shannon entropy – and the radius of the ball describes the *quantum* capacity of the quantum channel [12,39,51].

The classical capacity $C(\mathcal{N})$ of a quantum channel $\mathcal{N}$ is measured by the number of classical bits that can be transmitted per channel use. Although, in this case, the quantum mutual information is used to describe the channel capacity, in the case of the quantum capacity $Q(\mathcal{N})$ of the same quantum channel $\mathcal{N}$, we have to use the *quantum coherent information* $I_{coh} = I(\rho_A : \mathcal{N}(\rho_A))$, whose maximum is equal to $\max_{\rho_A}(\mathrm{S}(\sigma_B) - \mathrm{S}(\sigma_E))$, where the input of the channel is denoted by $\rho_A$, while $\mathrm{S}(\sigma_B)$ is the von Neumann entropy of Bob's state $\sigma_B = \mathcal{N}(\rho_A)$ and the information leaked to the environment is measured by $\mathrm{S}(\sigma_E)$. The *von Neumann entropy* of a density matrix $\rho$ is defined as $\mathrm{S}(\rho) = -Tr(\rho \log \rho)$.

$Q(\mathcal{N})$ describes the quantum capacity of the channel as the number of quantum bits per channel use that can be transmitted in a coherent state through a noisy quantum channel.

## 3.1 Quantum Capacity of a Quantum Channel

Holevo introduced the concept of the Holevo quantity of information [8,35,48,49]. A very important but not well-known fact was shown by Schumacher and Westmoreland in [48]: the quantum coherent information can be computed as the difference between two Holevo quantities – Holevo information $\mathcal{X}_{AB}$, which measures the Holevo quantity between Alice and Bob [48], and Holevo information $\mathcal{X}_{AE}$, which measures the Holevo quantity between Alice and the environment during the transmission of the quantum state [48]. In Fig. 4, we summarize a very important connection between the classical Holevo information and the quantum coherent information.

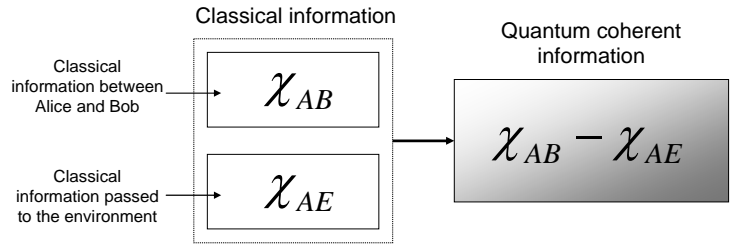

**Fig. 4.** Expression of the quantum coherent information in terms of Holevo quantities.

As follows, the quantum capacity can also be expressed as the difference between the von Neumann entropies of two channel output states. The first state is received by Bob, while the second one is received by a non-valid receiver – the environment [48]. Quantum coherent information plays a fundamental role in the superactivation of the quantum channel and, like the Holevo quantity in the classical *HSW (Holevo-Schumacher-Westmoreland)* capacity [8, 49] of a quantum channel, the quantum coherent information plays a crucial role in the asymptotic *LSD (Lloyd-Shor-Devetak)* capacity of a quantum channel [12,39,51]. To define the asymptotic LSD quantum capacity, we have to regularize the maximum of the quantum coherent information $I(\rho_A : \mathcal{N}(\rho_A))$, employing the parallel use of *n* copies of channel $\mathcal{N}$ as follows:

$$\begin{aligned} Q(\mathcal{N}) &= \lim_{n \to \infty} \frac{1}{n} Q^{(1)}\left(\mathcal{N}^{\otimes n}\right) \\ &= \lim_{n \to \infty} \frac{1}{n} \max_{\rho_A} I\left(\rho_A : \mathcal{N}^{\otimes n}(\rho_A)\right), \end{aligned} \tag{6}$$

where $Q^{(1)}(\mathcal{N}) = \max_{\rho_A} I(\rho_A : \mathcal{N}(\rho_A))$ is the *single use* quantum capacity of the quantum channel $\mathcal{N}$. In our paper, we use the fact that the *asymptotic* (i.e. not the single use) quantum capacity can be computed by the Holevo quantities as follows:

$$Q(\mathcal{N}) = \lim_{n \to \infty} \frac{1}{n} \max_{A^{\otimes n}, \rho_x^A} (\mathcal{X}_{AB} - \mathcal{X}_{AE}), \tag{7}$$

where

$$\mathcal{X}_{AB} = S(\mathcal{N}_{AB}(\sigma_{AB})) - \sum_i p_i S(\mathcal{N}_{AB}(\rho_i)) \tag{8}$$

and

$$\mathcal{X}_{AE} = S(\mathcal{N}_{AE}(\sigma_{AE})) - \sum_i p_i S(\mathcal{N}_{AE}(\rho_i)) \tag{9}$$

measure the Holevo quantities between Alice and Bob and between Alice and environment $E$, where $\sigma_{AB} = \sum_i p_i \rho_i$ and $\sigma_{AE} = \sum_i p_i \rho_i$ are the averaged states [48,49].

### 3.2 Quantum Relative Entropy Interpretation of Quantum Capacity

We use the results of Schumacher and Westmoreland [48,49] to describe geometrically the channel capacity of quantum channels in terms of the quantum relative entropy. They have shown that, for a given quantum channel $\mathcal{N}$, the Holevo quantity for every optimal output state $\rho_k$ and the $\sigma$ average state can be expressed as:

$$\mathcal{X}(\mathcal{N}) = D(\rho_k \| \sigma), \tag{10}$$

where $\sigma = \sum p_k \rho_k$ is the optimal average output state, and $D$, the relative entropy function of two density matrices, is defined as:

$$D(\rho_k \| \sigma) = Tr[\rho_k \log(\rho_k) - \rho_k \log(\sigma)]. \tag{11}$$

For non-optimal output states $\delta$ and optimal average output state $\sigma = \sum p_k \rho_k$, we have $\mathcal{X}(\mathcal{N}) = D(\delta \| \sigma) \leq D(\rho_k \| \sigma)$. Schumacher and Westmoreland [49] also showed that there is at least one optimal output state $\{p_k, \rho_k\}$ that achieves the optimum Holevo quantity $\mathcal{X}(\mathcal{N}) = D(\rho_k \| \sigma)$. The geometric interpretation of quantum channel capacity was introduced in [48] as the radius of the smallest ball inside the Bloch sphere that contains all channel output states, where "smallest" means, when using the quantum relative entropy function as a distance measure:

$$\mathcal{X}(\mathcal{N}) = r^* = \min_{\{\sigma\}} \max_{\{\rho\}} D(\mathcal{N}(\rho) \| \mathcal{N}(\sigma)). \tag{12}$$

If we denote the convex hull of possible channel output states [8] for channel $\mathcal{N}$ by $\mathcal{A}$ and the convex hull of the set of states by $\mathcal{B}$, then for $\mathcal{A} \in \rho$ and $\mathcal{B} \in \sigma$:

$$\mathcal{X}(\mathcal{N}) = r^* = \min_{\{\sigma\}} \max_{\{\rho\}} D(\rho \| \sigma). \tag{13}$$

Schumacher and Westmoreland [49] also proved that there is an optimum output state $\{p_k, \rho_k\}$ for every $\sigma$ that satisfies the maximization such that $\sigma = \sum p_k \rho_k$, and that the average output state $\sigma = \sum p_k \rho_k$, which maximizes the capacity for any optimal set of output states $\rho = \sum \{p_k, \rho_k\}$, is unique. We analyze the superactivation of the quantum channel by clustering and convex hull calculations based on the quantum relative entropy. If we denote the optimal output states that achieve the Holevo capacity $\mathcal{X}(\mathcal{N})$ of channel $\mathcal{N}$ by $\{\mathcal{N}(\psi_k) = p_k, \rho_k\}$ and the average is $\sigma = \sum_k p_k \rho_k$, then the quantum channel capacity can be derived in terms of the quantum relative entropy as follows [8,49]:

$$\begin{aligned}
\sum_k p_k D(\rho_k \| \sigma) &= \sum_k \left( p_k Tr[\rho_k \log(\rho_k)] - p_k Tr[\rho_k \log(\sigma)] \right) \\
&= \sum_k \left( p_k Tr[\rho_k \log(\rho_k)] \right) - Tr\left[ \sum_k (p_k \rho_k \log(\sigma)) \right] \\
&= \sum_k \left( p_k Tr[\rho_k \log(\rho_k)] \right) - Tr[\sigma \log(\sigma)] \\
&= \mathrm{S}(\sigma) - \sum_k p_k \mathrm{S}(\rho_k) = \mathcal{X}.
\end{aligned} \tag{14}$$

We will use the geometric interpretation of the Holevo quantity to express the asymptotic quantum capacity of a quantum channel. The geometric interpretation of the Holevo quantity has been studied by Cortese [8], who also extended these results to general qudit (i.e., higher dimensional) channels.

### 3.3 The Asymptotic Quantum Capacity

In our work, we use the results of Lloyd-Shor-Devetak [12,39,51] to analyze the superactivation property of quantum channels. According to the LSD theorem and the result of Schumacher and Westmoreland [49], the *single use* quantum capacity $Q^{(1)}(\mathcal{N})$ of a quantum channel $\mathcal{N}$ can be defined as the radius $r^*$ of the smallest quantum informational ball:

$$\begin{aligned} r^* = Q^{(1)}(\mathcal{N}) &= \max_{\{\text{all possible } p_i \text{ and } \rho_i\}} \mathcal{X}_{AB} - \mathcal{X}_{AE} \\ &= \max_{p_1,\ldots,p_n,\rho_1,\ldots,\rho_n} S\left(\mathcal{N}_{AB}\left(\sum_{i=1}^n p_i(\rho_i)\right)\right) - \sum_{i=1}^n p_i S(\mathcal{N}_{AB}(\rho_i)) \\ &\quad - S\left(\mathcal{N}_{AE}\left(\sum_{i=1}^n p_i(\rho_i)\right)\right) + \sum_{i=1}^n p_i S(\mathcal{N}_{AE}(\rho_i)), \end{aligned} \qquad (15)$$

where $\mathcal{X}_{AB}$ is the *Holevo quantity* of Bob's output, $\mathcal{X}_{AE}$ is the information leaked to the environment during the transmission, and $\mathcal{N}(\rho_i)$ represents the $i$-th output density matrix obtained from the quantum channel input density matrix $\rho_i$. Using the resulting quantum relative entropy function and the LSD theorem, the asymptotic quantum channel can be expressed with the help of the radii of the smallest quantum informational balls as follows:

$$\begin{aligned} Q(\mathcal{N}) &= \lim_{n\to\infty} \frac{1}{n} Q^{(1)}(\mathcal{N}^{\otimes n}) = \lim_{n\to\infty} \frac{1}{n}\left(\sum_{i=1}^n r_i^*\right) \\ &= \lim_{n\to\infty} \frac{1}{n} \max_{p_1,\ldots,p_n,\rho_1,\ldots,\rho_n} I_{coh} = \lim_{n\to\infty} \frac{1}{n} \max_{p_1,\ldots,p_n,\rho_1,\ldots,\rho_n} (\mathcal{X}_{AB} - \mathcal{X}_{AE}) \\ &= \lim_{n\to\infty} \frac{1}{n} \sum_n \left(\min_{\sigma_{1\ldots n}} \max_{\rho_{1\ldots n}} D(\rho_k^{AB} \| \sigma^{AB}) - D(\rho_k^{AE} \| \sigma^{AE})\right) \\ &= \lim_{n\to\infty} \frac{1}{n} \sum_n \left(\min_{\sigma_{1\ldots n}} \max_{\rho_{1\ldots n}} D(\rho_k^{AB-AE} \| \sigma^{AB-AE})\right), \end{aligned} \qquad (16)$$

where $r_i^*$ is the single use capacity of the $i$-th channel use of quantum channel $\mathcal{N}$, $\rho_k^{AB-AE}$ is the optimal output channel state, and $\sigma^{AB-AE}$ is the average state. The superscript *AB-AE* denotes the information that is transmitted from Alice to Bob minus the information that is leaked to the environment during the transmission. Using this result, the superactivation of the asymptotic quantum capacity [52] can be studied by using the quantum relative entropy function as a distance measure [4,10,17,42,56].

## 4. Geometrical Interpretation of Quantum Capacity

The relative entropy in classical systems is a measure of how close a probability distribution $p$ is to a model probability distribution $q$ [5,35] and is given by $D(p\|q) = \sum_i p_i \log \frac{p_i}{q_i}$, while the relative entropy between quantum states is measured by

$$D(\rho\|\sigma) = Tr\big[\rho\big(\log\rho - \log\sigma\big)\big]. \tag{17}$$

The quantum informational distance has some distance-like properties: $D(\rho\|\sigma) \geq 0$ iff $\rho \neq \sigma$, and $D(\rho\|\sigma) = 0$ iff $\rho = \sigma$. However, it is *not* symmetric [8,48]. The quantum relative entropy between a general quantum state $\rho = (x, y, z)$ and a mixed state $\sigma = (\tilde{x}, \tilde{y}, \tilde{z})$ with radii $r_\rho = \sqrt{x^2 + y^2 + z^2}$, and $r_\sigma = \sqrt{\tilde{x}^2 + \tilde{y}^2 + \tilde{z}^2}$ is given by:

$$D(\rho\|\sigma) = \frac{1}{2}\log\frac{1}{4}\big(1 - r_\rho^2\big) + \frac{1}{2}r_\rho \log\frac{(1+r_\rho)}{(1-r_\rho)} - \frac{1}{2}\log\frac{1}{4}\big(1 - r_\sigma^2\big) - \frac{1}{2r_\sigma}\log\frac{(1+r_\sigma)}{(1-r_\sigma)}\langle\rho,\sigma\rangle, \tag{18}$$

where $\langle\rho,\sigma\rangle = (x\tilde{x} + y\tilde{y} + z\tilde{z})$. For a maximally mixed state $\sigma = (\tilde{x}, \tilde{y}, \tilde{z}) = (0, 0, 0)$ and $r_\sigma = 0$, the quantum relative entropy can be expressed as:

$$D(\rho\|\sigma) = \frac{1}{2}\log\frac{1}{4}\big(1 - r_\rho^2\big) + \frac{1}{2}r_\rho \log\frac{(1+r_\rho)}{(1-r_\rho)} - \frac{1}{2}\log\frac{1}{4}. \tag{19}$$

The relative entropy of quantum states can be described by a strictly convex and differentiable generator function $\mathbf{F}$:

$$\mathbf{F}(\rho) = -\mathrm{S}(\rho) = Tr(\rho \log \rho), \tag{20}$$

where $-\mathrm{S}$ is the negative entropy of quantum states. The quantum relative entropy $D(\rho\|\sigma)$ for density matrices $\rho$ and $\sigma$ is given by $\mathbf{F}$ as follows:

$$D(\rho\|\sigma) = \mathbf{F}(\rho) - \mathbf{F}(\sigma) - \langle\rho - \sigma, \nabla\mathbf{F}(\sigma)\rangle, \tag{21}$$

where $\langle\rho,\sigma\rangle = Tr(\rho\sigma^*)$ is the inner product of quantum states and $\nabla\mathbf{F}(\cdot)$ is the gradient.

In Fig. 5, the quantum informational distance, $D(\rho\|\sigma)$, is the vertical distance between the generator function $\mathbf{F}$ and $H(\sigma)$, the hyperplane tangent to $\mathbf{F}$ at $\sigma$ [44,45]. The point of intersection of quantum state $\rho$ on $H(\sigma)$ is denoted by $H_\sigma(\rho)$.

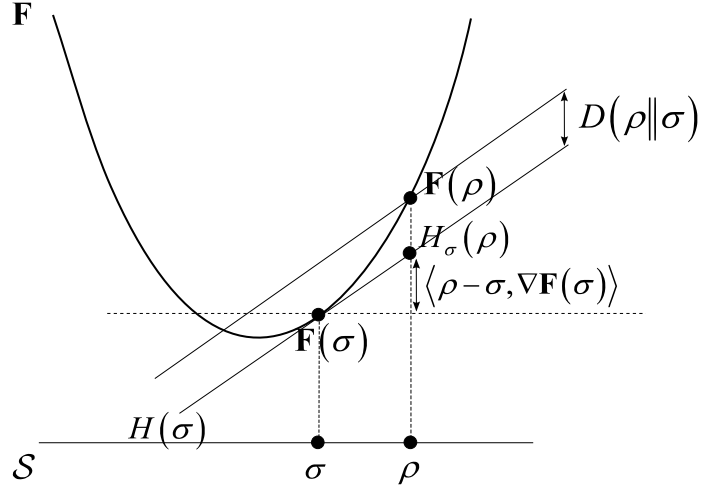

**Fig. 5.** Depiction of generator function as a negative von Neumann entropy.

For the quantum informational distance function, the generator function is the negative von Neumann entropy function $-\mathrm{S}$,

$$\mathbf{F}(\rho) = -\mathrm{S}(\rho) = Tr(\rho \log \rho), \tag{22}$$

where $F : S(\mathbb{C}^d) \to \mathbb{R}$. The quantum informational distance function $D_\mathbf{F}(\rho \| \sigma)$ with generator function $\mathbf{F}(\rho) = -\mathrm{S}(\rho)$ is illustrated in Fig. 6. The generator function of the quantum informational distance is the negative von Neumann entropy function.

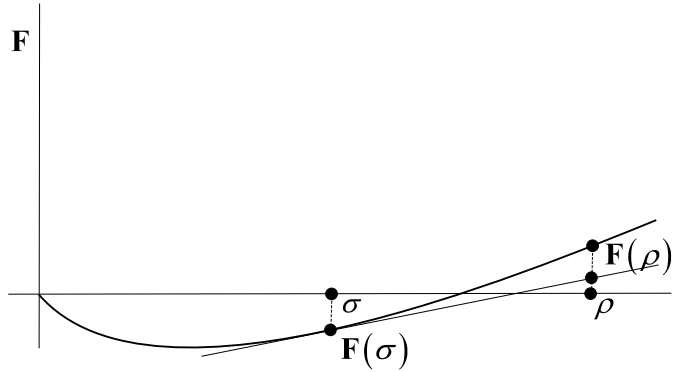

**Fig. 6.** Negative von Neumann generator function.

The quantum informational distance function is linear; thus, for convex functions $\forall \mathbf{F}_1 \in \mathcal{C}$ and $\forall \mathbf{F}_2 \in \mathcal{C}$, $D_{\mathbf{F}_1 + \lambda \mathbf{F}_2}(\rho \| \sigma) = D_{\mathbf{F}_1}(\rho \| \sigma) + \lambda D_{\mathbf{F}_2}(\rho \| \sigma)$ for any $\lambda \geq 0$. The geometric structure of quantum informational balls differs from the geometric structure of ordinary Euclidean balls. In Fig. 7, we have illustrated the circumcenter $\mathbf{c}^*$ of $\mathcal{S}$ for the Euclidean distance and for quantum relative entropy [44,45]. For a triangle, the circumcenter is defined as the center of a circle that circumscribes the triangle [1,3,7].

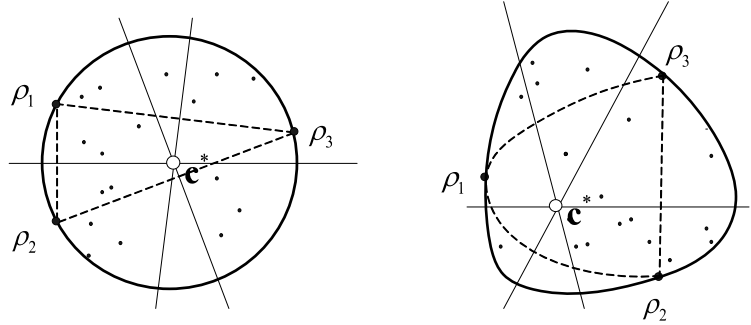

**Fig. 7.** Circumcenter for Euclidean distance and quantum relative entropy.

In Fig. 8, we compare the smallest quantum informational ball and the ordinary Euclidean ball.

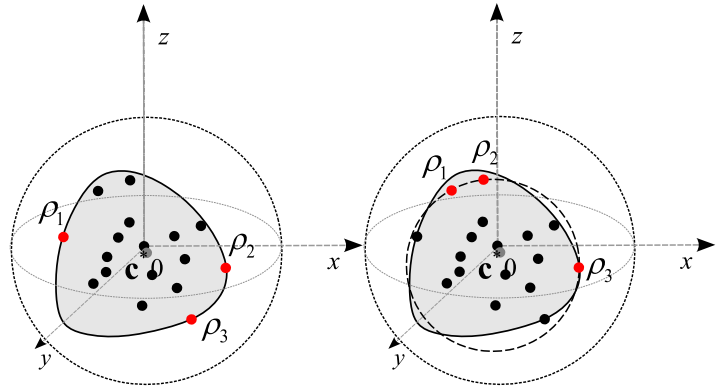

**Fig. 8.** The smallest balls differ for the quantum informational distance and Euclidean distance.

Thus, the quantum states $\rho_1$, $\rho_2$, and $\rho_3$, which determine the smallest enclosing ball in Euclidean geometry, differ from the states that would do the same for the quantum informational ball.

**4.1 Asymptotic Quantum Capacity and Superball Representation**

In this section, we define a new geometric object to describe the asymptotic quantum capacity of the quantum channel. According to the LSD theorem [12,39,51], using the optimal output state $\rho_k^{AB-AE}$ and channel average state $\sigma^{AB-AE}$ of $n$ uses of the quantum channel $\mathcal{N}$, the radius $r_{super}^*$ of the smallest enclosing quantum informational *superball* can be expressed as:

$$\begin{aligned} r^*_{super}(\mathcal{N}) &= Q(\mathcal{N}) = \lim_{n \to \infty} \frac{1}{n} Q^{(1)}(\mathcal{N}^{\otimes n}) \\ &= \lim_{n \to \infty} \frac{1}{n} \sum_n \left( \min_{\sigma_{1...n}} \max_{\rho_{1...n}} D\left( \rho_k^{AB-AE} \middle\| \sigma^{AB-AE} \right) \right). \end{aligned} \qquad (23)$$

In our geometric representation, we analyze the superactivation property of the quantum channel using the *mini-max* criterion for states $\rho_k^{AB-AE}$ and $\sigma^{AB-AE}$. The radius $r^*_{super}$ of the superball, measured using the relative entropy function as the distance measure [48], is equal to the asymptotic quantum capacity. Using the radius $r^*$ of the quantum informational ball to express the *single use* channel capacity, the *asymptotic* quantum capacity $Q(\mathcal{N}) = \lim_{n \to \infty} \frac{1}{n} Q^{(1)}(\mathcal{N}^{\otimes n})$ of a quantum channel $\mathcal{N}$ can be expressed as:

$$r^*_{super}(\mathcal{N}) = Q(\mathcal{N}) = \lim_{n \to \infty} \frac{1}{n} \left( \sum_{i=1}^{n} r_i^* \right), \qquad (24)$$

where $r_i^*$ is the radius of the smallest quantum informational ball, which describes the *single use* quantum capacity of the *i*-th use of the quantum channel [12,39,51].

In the superactivation problem, we have to use different quantum channel models $\mathcal{N}_1$ and $\mathcal{N}_2$ [9,52]. For two quantum channels $\mathcal{N}_1$ and $\mathcal{N}_2$, the $Q(\mathcal{N}_1 \otimes \mathcal{N}_2)$ *asymptotic* quantum channel capacity of the joint structure can be expressed by superball radius

$$\begin{aligned} r^*_{super}(\mathcal{N}_1 \otimes \mathcal{N}_2) &= Q(\mathcal{N}_1 \otimes \mathcal{N}_2) \\ &= \lim_{n \to \infty} \frac{1}{n} Q^{(1)}\left( (\mathcal{N}_1 \otimes \mathcal{N}_2)^{\otimes n} \right). \end{aligned} \qquad (25)$$

As we have depicted in Fig. 9, the quantum information superball differs from the classical Euclidean ball: in comparison, it has a *distorted* geometrical structure [21,44,45]. The radius of the quantum informational superball describes the $Q(\mathcal{N}_1 \otimes \mathcal{N}_2)$ asymptotic quantum capacity of the $\mathcal{N}_1 \otimes \mathcal{N}_2$ joint channel structure.

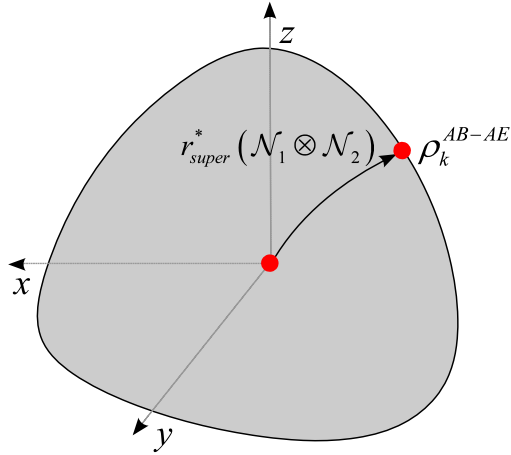

**Fig. 9.** The quantum superball defined for the analysis of superactivation of zero-capacity quantum channels.

In Fig. 10, we summarize our geometrical iteration process as follows. The inputs of the geometric construction of the quantum superball are the channel output states of the two separate quantum channels $\mathcal{N}_1^{\otimes n}$ and $\mathcal{N}_2^{\otimes n}$. This process yields the superball radius as described above. The iterated joint channel construction is denoted by $(\mathcal{N}_1 \otimes \mathcal{N}_2)^{\otimes n}$, while the output of the rounded box is the radius $r^*_{super}(\mathcal{N}_1 \otimes \mathcal{N}_2)$ of the quantum informational superball.

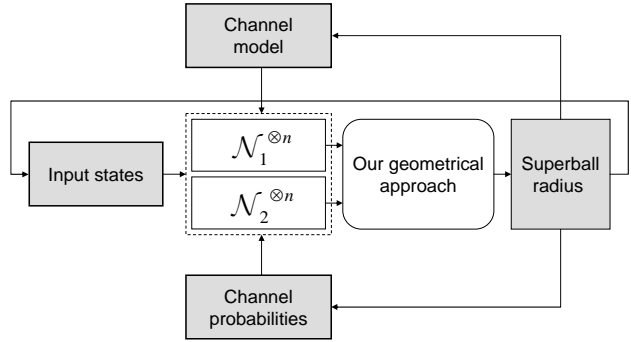

**Fig. 10.** The recursive algorithm iterates on the input, channel models, and error probabilities of the channels to find a combination for which superactivation holds.

The recursive iterations are made on three parameters: quantum channel models, channel probabilities, and set of input states. According to the length of the superball radius, the iteration stops if the conditions for superactivation hold.

# 5. Efficient Coreset Construction of Channel Output States

The coreset technique has deep relevance to classical computational geometry. A coreset of a set of output quantum states has the same behavior as the larger input set, so clustering and other approximations can be made with smaller coresets. The coreset can be viewed as a smaller input set of channel output states; hence, it can be used as the input to an approximation algorithm. The weighted sum of errors of the smaller coreset is a $(1 \pm \varepsilon)$-approximation of the larger input set. The bound on this error can be decreased only if the center points that form a finite set are used in the approximation. These coresets are called weak coresets [15], and this method can be applied in quantum space between quantum states. Using weak coresets, the run time of $(1 + \varepsilon)$ coreset algorithms [15,31] with respect to the quantum informational distance can be improved. To construct the coreset method analyzing the superactivation of zero-capacity quantum channels [20], we have to introduce the definition of *similar quantum informational distances* and *weak coresets of quantum states*.

## 5.1 Similar Quantum Informational Distance

The quantum informational distance is asymmetric and contains singularities since there are density matrices $\rho$ and $\sigma$ for which $D(\rho, \sigma) = \infty$. The similar quantum informational divergence function does not contain these singularities, and the distances are approximately symmetric. To use a similar quantum informational divergence function, we first define it as follows. The quantum informational distance function $D(\rho \| \sigma)$ between density matrices $\rho$ and $\sigma$ is $\mu$-similar for a positive real constant $\mu$ if there is a positive definite matrix $A$ such that:

$$\mu D_A(\rho \| \sigma) \leq D(\rho \| \sigma) \leq D_A(\rho \| \sigma). \tag{26}$$

For quantum informational distances, if the domain is given as $\mathcal{D} = [\lambda, \gamma] \subseteq R_+^d$, then $\mu = \frac{\lambda}{\gamma}$ and $\frac{1}{2\lambda}I$. If we have $0 < \lambda < \gamma$, then the quantum informational distance function can be calculated as $D(\rho, \sigma) = Tr[\rho(\log \rho - \log \sigma)] - \sum(\rho_i - \sigma_i)$ on the domain $\mathcal{D} = [\lambda, \gamma] \subseteq R^d$

[1]. The quantum informational distance is $\mu$-similar if $\mu = \frac{\lambda}{\gamma}$ and $A = \frac{1}{2\lambda}I$. In these cases, the quantum informational distance function is $\mu$-similar because it is restricted to a sub-domain, which *avoids the singularities* [1], [20]. It can be easily proven that the quantum informational distance function is strictly convex and that all second-order partial derivates exist and are continuous on the domain $\mathcal{D} = [\lambda, \gamma] \subseteq R^d$ with parameters $\mu = \frac{\lambda}{\gamma}$ and $A = \frac{1}{2\lambda}I$ [1]. The applied coreset algorithm was originally presented by Chen [6,7]. We show that this method can be used to generate a coreset based on a similar quantum informational distance function. To obtain an enhanced version of previously known coreset approximation algorithms, we must define weak coresets.

### 5.2 Weak Coreset of Quantum States

Weak coresets include all the relevant information required to analyze the original extremely large input set [15]. The coreset approach has significantly lower computational complexity; hence, it can be applied very efficiently [31]. In our method, weak coresets are applied to $\mu$-similar quantum informational distances since, in these subsets, the distances between quantum states are symmetric; hence, singularities can be avoided and fast Euclidean methods can be applied [30,34,54].

The superactive quantum channels can be discovered with approximation error $(1 \pm \varepsilon)$ by using the smaller $\mu$-similar subset of input quantum states. Using the results from Chen's work [6,7] we show that, by using this algorithm, the superactivation of quantum channels can be approximated with error $(1 \pm \varepsilon)$ using the smaller $\mu$-similar subset of input quantum states. The goal of the algorithm is to find a set of size $k$ such that the sum of errors of quantum informational distances is minimized; hence, $error(\rho_i, \sigma) = \sum_{i=1}^{n} \min D_\sigma(\rho_i \| \sigma)$. The algorithm solves the *k*-median problem with respect to the quantum informational distance $D_\sigma$ in quantum space [20]. The output of the algorithm is a set of $k$ quantum states for which the function $error(\rho_i, \sigma)$ is minimized. We generalize the *k*-median problem for quantum informational distances. Let us assume that we have two quantum states $\rho$ and $\sigma$ in domain $\mathcal{S}$. We would like to construct a subset of $\mathcal{S}_{OUT}$ of $k$ quantum states, for which:

$$D(\rho, \mathcal{S}_{OUT}) = \min_{\sigma \in \mathcal{S}^*} D(\rho \| \sigma). \tag{27}$$

The $k$-median problem for quantum states can be stated as follows. We would like to use only a finite set $\mathcal{S}_{IN}$ of quantum states from the original larger space. For a set $\mathcal{S}_{IN}$, we would like to construct a set $\mathcal{S}_{OUT}$ of $k$-quantum states, for which $error(\mathcal{S}_{IN}, \mathcal{S}_{OUT}) = \sum_{\rho \in \mathcal{S}_{IN}} D_\sigma(\rho \| \mathcal{S}_{OUT})$ is minimized; hence [20]:

$$error(\mathcal{S}_{IN}, \mathcal{S}^*) = \sum_{\rho \in \mathcal{S}_{IN}} \min D_\sigma(\rho \| \mathcal{S}^*). \tag{28}$$

The error of the optimal solution for input states $\mathcal{S}_{IN}$ is denoted by $opt_k(\mathcal{S}_{IN})$, and the elements of the output set $\mathcal{S}_{OUT}$ are the $k$ median-quantum states of set $\mathcal{S}_{IN}$. To construct a more efficient algorithm, we use only the $\mu$-similar quantum informational distances; hence, the set of input quantum states $\mathcal{S}_{IN}$ is restricted to quantum states for which the singularities can be avoided [1].

The superactivation properties of quantum channels can be discovered by using $\mu$-similar quantum informational distances and the coreset construction method. For any set $\mathcal{S}_{IN}$ of size $n$ quantum states and for any finite $\mathcal{W} \subseteq \mathcal{S}$, there is a weak coreset of size $\mathcal{O}\left(\frac{1}{\varepsilon^2} k \log(n) \log\left(k |\mathcal{W}|^k \log n\right)\right)$. This $\mathcal{W}$-weak coreset of quantum states can be constructed in time $\mathcal{O}\left(\frac{1}{\varepsilon^2} k \log(n) \log\left(k |\mathcal{W}|^k \log n\right) + dkn\right)$, where $k$ is the number of quantum states in set $\mathcal{S}_{OUT}$, $n$ is the number of input states, and $d$ is the dimension of the points. The previous result can be integrated into our analysis as follows. Using $\mu$-similar quantum informational distances and the $\mathcal{W}$-weak coreset of quantum states, the superactivation of quantum channels can be analyzed by an $(1+\varepsilon)$-approximation algorithm in a run time $\mathcal{O}\left(d^2 2^{\frac{k}{\varepsilon}} \log^{k+2} n + dkn\right)$.

Using the result of Banerjee et al. [3], the optimal 1-median of any given input set $\mathcal{S}$ in quantum space can be uniquely defined by the centroid $c = \frac{1}{|\mathcal{S}|}\sum_{\rho \in \mathcal{S}} \rho$. Using the fact that an optimal solution of the $k$-median clustering problem can be approached by $(k-1)$ linearly separable subsets, it can be shown that, for any set $\mathcal{S}_{IN}$, most $n^{dk}$ states have to be considered optimal $k$-median quantum states of $\mathcal{S}_{IN}$ [1,15]. We use a smaller set $\mathcal{S}$ from $\mathcal{S}_{IN}$, which is a small weighted set that has the same clustering behavior as the larger input set $\mathcal{S}_{IN}$. The coreset method used in our approach can be defined by the error of the approximation in terms of the quantum informational distance between quantum states as follows [20]:

$$error_w(\mathcal{S}, \mathcal{S}_{OUT}) = \sum_{\rho \in \mathcal{S}} w(\rho) D(\rho \| \mathcal{S}_{OUT}), \tag{29}$$

and this error is a $(1 \pm \varepsilon)$-approximation of $error(\mathcal{S}_{IN}, \mathcal{S}_{OUT})$ for any set of quantum states $\mathcal{S}_{OUT}$ of size $|\mathcal{S}_{OUT}| = k$. For the weak coreset construction, let us assume that we have a set of quantum states $\mathcal{S}_{IN}$ and a set $\mathcal{W}$. If the weight function is defined by:

$$\sum_{\mathcal{S}} w(\rho) = |\mathcal{S}_{IN}|, \tag{30}$$

then the weighted set $\mathcal{S}$ is a $\mathcal{W}$-weak coreset of $\mathcal{S}_{IN}$, iff for all $\mathcal{S}_{OUT} \in \mathcal{W}$ of size $|\mathcal{S}_{OUT}| = k$; we have:

$$|error(\mathcal{S}_{IN}, \mathcal{S}_{OUT}) - error_w(\mathcal{S}, \mathcal{S}_{OUT})| \leq (\varepsilon) error(\mathcal{S}_{IN}, \mathcal{S}_{OUT}). \tag{31}$$

This $\mathcal{W}$-weak coreset is called the $(k, \varepsilon)$ weak-coreset of $\mathcal{S}_{IN}$. To get this construction with this error bound, we use the results of Chen [6].

*5.2.1 Coreset Method for Quantum Informational Distances*

To apply the modified coreset method, we have to construct a $[\alpha, \beta]$ bicriteria approximation algorithm to get the set of median quantum states $M = \{\sigma_1, \sigma_2, \ldots \sigma_k\}$ of a $k$-median clustering of $\mathcal{S}_{IN}$, for which $error(\mathcal{S}_{IN}, M) \leq \alpha opt_k(\mathcal{S}_{IN})$ and $|M| = k \leq \beta k$. Using the results of [1,31] the bicriteria algorithm can be summarized as follows [20]:

> **Bicriteria algorithm to channel analysis**
> 1. Choose an initial quantum state $\sigma_1$ uniformly at random from $\mathcal{S}_{IN}$
> 2. Let $M$ be the set of chosen quantum states from $\mathcal{S}_{IN}$. State $\rho \in \mathcal{S}_{IN}$ is chosen with probability $\dfrac{D(\rho\|M)}{error(\mathcal{S}_{IN}, M)}$ as the next state of $M$.
> 3. Repeat step 2 until $M$ contains $k$ quantum states.

At the end of the bicriteria algorithm, we have a set of median quantum states $M = \{\sigma_1, \sigma_2, \dots \sigma_k\}$, for which $error(\mathcal{S}_{IN}, M) \leq \alpha opt_k(\mathcal{S}_{IN})$ and $|M| = k \leq \beta k$. After application of the bicriteria algorithm, we use the modified coreset construction method presented in [6] for the quantum states as follows [20]:

> **Coreset algorithm to quantum channel analysis**
> 1. Partition $\mathcal{S}_{IN}$ into $\mathcal{S}_1, \mathcal{S}_2 \dots, \mathcal{S}_k$ by assuming each quantum state $\rho \in \mathcal{S}_{IN}$ to their closest $\sigma_i \in M$.
> 2. Let $\rho \in \mathcal{S}_{IN}$ iff $\sigma_i = \arg\min_{\sigma \in M} D(\rho\|\sigma)$.
> 3. Let $R = \dfrac{1}{\alpha n} error(\mathcal{S}_{IN}, M)$.
> 4. Define quantum informational ball $\mathcal{B}(\sigma_i)$ with radius $r^*$ and center $\sigma_i$ as follows:
>    $\mathcal{B}(\sigma_i) = D(x\|\sigma_i) \leq r$.
> 5. Define the partition of $\{\mathcal{S}_{ij}\}_{i,j}$ of $\mathcal{S}_{IN}$ by $\mathcal{S}_{ij} = \mathcal{S}_i \cap \mathcal{B}(\sigma_i)$ for $i = 1,2,\dots,k$.
> 6. Let $\mathcal{S}_{ij} = \mathcal{S}_i \cap (\mathcal{B}_{2^j}(\sigma_i) \setminus \mathcal{B}_{2^{j-1}}(\sigma_i))$ for $i = 1,2,\dots,k$ and $j = i = 1,2,\dots \gamma$, where $\gamma = \lceil \log(\alpha n) \rceil$.
> 7. For $i,j$ let $\mathcal{S}_{ij}$ be a uniform set from $\mathcal{S}_{IN}$ of size $|\mathcal{S}_{ij}| = m$.
> 8. Let $w(\rho) = \dfrac{1}{m}|\mathcal{S}_{ij}|$ be the weight associated with $\rho \in \mathcal{S}_{ij}$.
> 9. Define the *weak coreset* $\mathcal{S}$ of input quantum states $\mathcal{S}_{IN}$ as follows:
>    $\mathcal{S} = \bigcup_{i,j} \mathcal{S}_{ij}$ of size $|\mathcal{S}| = mk\gamma = m\beta k \lceil \log(\alpha n) \rceil$.

The algorithm has approximation error $(1 + \varepsilon)$; however, the run time of the proposed method is more efficient since it uses the $\mathcal{W}$-weak coreset of set $\mathcal{S}_{IN}$ generated by Chen's algorithm instead of the original input set $\mathcal{S}_{IN}$.

### 5.3 Determination of Median-Quantum States

The clustering method of [1] for a weak set of quantum states and $\mu$-similar quantum informational distances can be summarized as follows [20]:

---

**CLUSTER** : **Clustering of channel output states**

1. Let $\mathcal{S}_{IN}$ be the set of remaining input states, with $w(\mathcal{S}_{IN}) = n$
2. Let $w$ be the weight function on input quantum states $\mathcal{S}_{IN}$
3. Let $m$ be the number of median-quantum states *yet to be* found
4. Let $C$ be the set of medians *already* found
5. **if** $m = 0$ then return $C$
6. **else**
7.     **if** $m \geq |\mathcal{S}_{IN}|$ then return $C \cup \mathcal{S}_{IN}$
8.     **else**
9.         Sample a multiset of quantum states $\mathcal{M}$ of size $\dfrac{96k^2}{\varepsilon^2 \mu \delta}$ from $\mathcal{S}_{IN}$
10.         $T \leftarrow$ Let $c$ the weighted centroid of $\mathcal{M}' \subseteq \mathcal{M}$, with $|\mathcal{M}'| = \dfrac{3}{\varepsilon \mu \delta}$
11.         **for** all $c \in T$ **do**
12.             $C^{(c)} \leftarrow$ **CLUSTER**$(\mathcal{S}_{IN}, w, m-1, C \cup \{c\})$
13.         **end for**
14.         Partition of the set of input quantum states $\mathcal{S}_{IN}$ into set $N$ and $\mathcal{S}_{IN} \setminus N$ such that:
15.         $\forall \rho \in N, \sigma \in \mathcal{S}_{IN} \setminus N : D(\rho \| C) \leq D(\sigma \| C)$ and
16.         $w(N) = w(\mathcal{S}_{IN} \setminus N) = \dfrac{n}{2}$
17.         Let $w^*$ the new weight function on $\mathcal{S}_{IN} \setminus N$
18.         Let $C^* \leftarrow$ **CLUSTER**$(\mathcal{S}_{IN} \setminus N, q, m, C)$
19.         **return** $C^{(c)}$ or $C^*$ with *minimum error*
20.     **end if**
21. **end if**

---

In Fig. 11, we illustrate the clustering of channel output states. In the clustering process, our algorithm computes the median-quantum states denoted by $\sigma_i$ using a fast weak coreset and clustering algorithm. In the next step, we compute the convex hull of the median quantum states and, from the convex hull, the radius of the smallest quantum informational ball can be obtained. The smallest superball measures the channel capacity; hence, the radius of the superball is equal to the sum of the radii of the quantum balls of independent channel outputs. The output states are measured by a joint measurement setting.

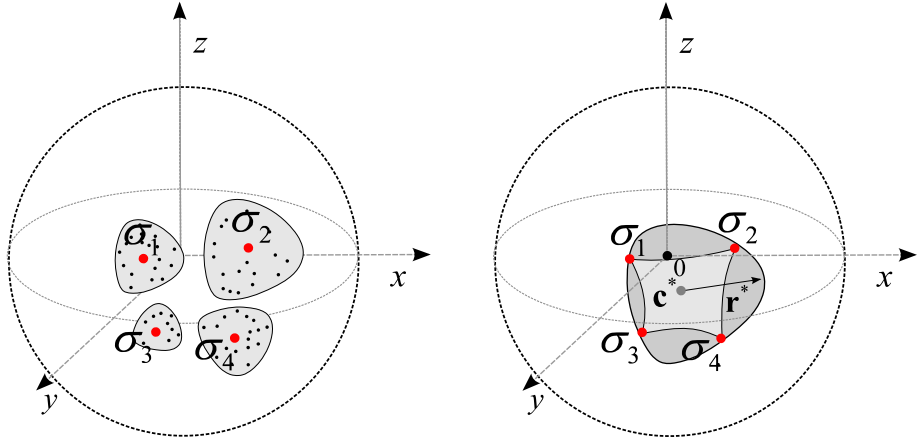

**Fig. 11.** Clustering of quantum states with the computed median quantum states.

To summarize, our quantum channel superactivation method combines the weak coreset method of Chen and the clustering algorithm presented by Ackermann et al. [1]. To use Chen's method [6,7] to construct the weak coreset, we apply a bicriteria algorithm [15] to find the required parameters, and then we apply the result of Ackermann et al. [1] in quantum space. In both methods, we use quantum informational distance functions as distance measures.

## 6. Illustrative Example

In this section, we show that our method can be used to confirm the results of [52,53] and that the method can be extended to a larger set of possible channels. Smith and Yard showed that any private Horodecki channel can be combined with other symmetric channels; hence, two zero-capacity quantum channels can be combined to realize a capacity greater than 0.01 [52]. The term "superactivation" describes this effect, which cannot be imagined for classical systems, i.e., it has no analogue in classical systems. Two zero-capacity quantum communication channels can be used to transmit information, and it is possible to "activate" one channel with the other channel so that the zero-capacity of the channels can be increased.

At present, we have no theoretical background to describe all possible combinations; hence, there should be many other possible combinations of superactive zero-capacity quantum channels. Our method provides a fundamentally new and efficient algorithmic solution to discover all possible "superactivated" channels, and it can be applied to analyze the capacity of these channels. Smith and Yard had left open the question as to whether how many different channel models can be used for superactivation [52]. We will also use the convexity of quantum channel capacity. The fact that quantum capacity is not a convex function of the channel can be used in our computations. There are zero-capacity quantum channels that can be combined to achieve higher capacity than the average capacity of the individual quantum channels; hence, for these channels, convexity does not hold.

We derive the joint channel capacities from the independent channel capacities, and we introduce a new representation, the superball representation, which was used for channel additivity analysis by Gyongyosi and Imre in [21]. In this paper, we describe only the clustering process of channel output quantum states to construct median-quantum states. The method of convex hull calculation from these median-quantum states is based on quantum Delaunay tessellation, a method proposed by Gyongyosi and Imre in [21]. In the introduced superball representation, the length of the superball radius is equal to the sum of the radii of the quantum informational balls of the independent channel capacities, measured in a joint measurement setting.

### 6.1 Confirmation of Result for Superactivation

In [52] the authors used a relationship between classical private and quantum channel capacities that could hold for any quantum channel $\mathcal{N}$. They considered a quantum channel for which $I(X:B) - I(X:E) \leq P^{(1)}(\mathcal{N})$, where $P^{(1)}(\mathcal{N})$ is the private capacity of channel $\mathcal{N}$. According to our method, the channel capacity is measured by the radius of the smallest quantum informational ball; we use the following equation to describe the private capacity of superactivated zero-capacity quantum channels:

$$r_{\mathcal{N}}^{private} = \max_{X} I(X:B) - I(X:E) = P^{(1)}(\mathcal{N}), \tag{32}$$

where $r_{\mathcal{N}}^{private}$ measures the single use classical private capacity of channel $\mathcal{N}$.

Every Horodecki channel $\mathcal{N}_H$ satisfies the relation $P(\mathcal{N}_H) > 0$; hence, there is an input for which $r_{\mathcal{N}_H}^{private} > 0$. The combination of a Horodecki channel $\mathcal{N}_H$ and a 50%-erasure channel $\mathcal{A}_e$-channel can result in the following superball radius:

$$r^* = Q^{(1)}(\mathcal{N}_H \otimes \mathcal{A}_e) = \frac{1}{2}(I(X:B) - I(X:E)) = \frac{1}{2}P^{(1)}(\mathcal{N}_H) = \frac{1}{2}r_{\mathcal{N}_H}^{private}, \tag{33}$$

while for the *superactivated asymptotic* joint quantum capacity with radius $r^{*}_{(\mathcal{N}_{H} \otimes \mathcal{A}_{e})}$ :

$$Q(\mathcal{N}_H \otimes \mathcal{A}_e) = r^{*}_{(\mathcal{N}_H \otimes \mathcal{A}_e)} \geq r^{*} = \frac{1}{2}P^{(1)}(\mathcal{N}_H) = \frac{1}{2}r^{private}_{\mathcal{N}_H}. \tag{34}$$

We show that our method finds that superactivated channel combination, with approximation error $\varepsilon$. The authors of [52] defined a private Horodecki channel $\mathcal{N}_H$ such that $r^{private}_{\mathcal{N}_H} = 0.02$, whose combination with $\mathcal{A}_e$ has a channel capacity $r^{*}_{(\mathcal{N}_H \otimes \mathcal{A}_e)} > \frac{1}{2}(0.02) = 0.01$. For the defined four-dimensional Horodecki channel, the following equation gives the private channel capacity:

$$r^{private}_{\mathcal{N}_H} \geq 1 - q\log q - (1-q)\log(1-q) > 0.02 \tag{35}$$

where $q = \frac{\sqrt{2}}{1+\sqrt{2}}$, a parameter used in the Kraus representation of the channel [52].

Here, we have used the fact that the map of any quantum channel can be written in Kraus form as $\mathcal{N}(\rho) = \sum_k N_k \rho N_k^{\dagger}$, where $N_k$ denotes the Kraus matrices, with $\sum_k N_k N_k^{\dagger} = I$. In the case of a Horodecki channel $\mathcal{N}_H$, the channel can be specified by six Kraus matrices [39,52]. We show that the results of [52] for the superactivation of a Horodecki channel $\mathcal{N}_H$ and an erasure channel $\mathcal{A}_e$, can be found and confirmed by our fundamentally new geometric approach, with approximation error $\varepsilon$. As we conclude, our method can be extended to other possible channel models and combinations of other channels and channel probabilities.

To describe geometrically the superactivation of zero-capacity quantum channels, we introduce the channel model and channel parameter $p$ as follows:

$$\mathcal{N} = p\mathcal{N}_H \otimes |0\rangle\langle 0| + (1-p)\mathcal{A}_e \otimes |1\rangle\langle 1|, \tag{36}$$

where $0 \leq p \leq 1$. The defined $\mathcal{N}$ channel model is the convex combination of two zero-capacity channels $\mathcal{N}_H \otimes |0\rangle\langle 0|$ and $\mathcal{A}_e \otimes |1\rangle\langle 1|$. Let us assume that we use these two quantum channels and their product channel representation $\mathcal{N}_1 \otimes \mathcal{N}_2$. The main goal of our geometric analysis is to find a channel probability parameter $p$ for which the joint capacity of the tensor product channel $\mathcal{N}_1 \otimes \mathcal{N}_2$ is greater than zero.

The channel construction technique for the superactivation of zero-capacity channels is illustrated in Fig. 12. To superactivate zero-capacity quantum channels, we must use the convex combination of different channel models and the *probabilistic mixtures* of these channels to realize superactivation.

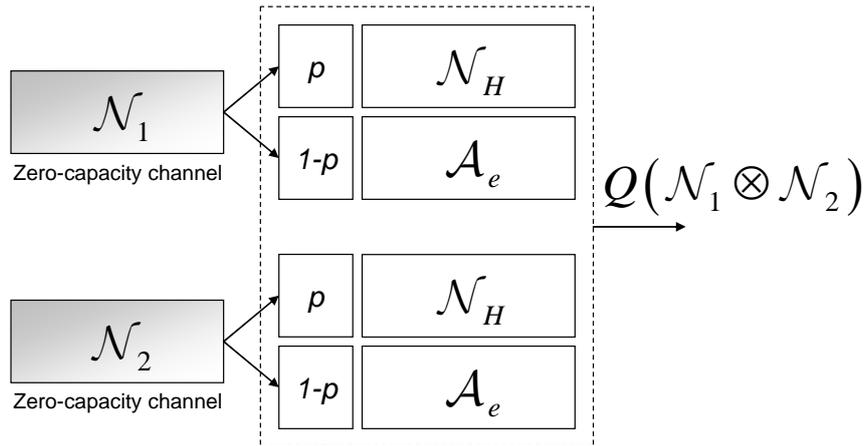

**Fig. 12.** Channel construction for the superactivation of zero channels.

In our iteration process, we search for the optimal channel probability parameter. In this case, we use the channel model of [52]; hence, we have fixed channel models, and we have to iterate on parameter $p$ only. The joint channel quantum capacity $Q(\mathcal{N}_1 \otimes \mathcal{N}_2)$ as radius $r^*_{super}(\mathcal{N}_1 \otimes \mathcal{N}_2)$ in function of channel probability $p$ can be described as follows:

$$r^*_{super}(\mathcal{N}_1 \otimes \mathcal{N}_2) = p^2 r^*_{(\mathcal{N}_H \otimes \mathcal{N}_H)} + p(1-p) r^*_{(\mathcal{N}_H \otimes \mathcal{A}_e)} \\ + (1-p) p r^*_{(\mathcal{A}_e \otimes \mathcal{N}_H)} + (1-p)^2 r^*_{(\mathcal{A}_e \otimes \mathcal{A}_e)}. \tag{37}$$

As shown in [52], the term $(1-p)^2 r^*_{(\mathcal{A}_e \otimes \mathcal{A}_e)}$ can be neglected since the quantum capacity of this combination is zero, i.e., $r^*_{(\mathcal{A}_e \otimes \mathcal{A}_e)} = 0$. As follows, the channel model is reduced to

$$Q(\mathcal{N}_1 \otimes \mathcal{N}_2) = r^*_{super}(\mathcal{N}_1 \otimes \mathcal{N}_2) = p^2 r^*_{(\mathcal{N}_H \otimes \mathcal{N}_H)} + 2p(1-p) r^*_{(\mathcal{N}_H \otimes \mathcal{A}_e)}, \tag{38}$$

and the radius of the smallest superballs can be described as

$$p^2 r^*_{(\mathcal{N}_H \otimes \mathcal{N}_H)} = p^2 Q^{(1)}(\mathcal{N}_H \otimes \mathcal{N}_H), \tag{39}$$

or

$$2p(1-p)r^*_{(\mathcal{N}_H \otimes \mathcal{A}_e)} = 2p(1-p)Q^{(1)}(\mathcal{N}_H \otimes \mathcal{A}_e), \tag{40}$$

where $0 < p < 1$. Here we note, the notation $\mathcal{N}_1 \otimes \mathcal{N}_2$ means using the joint channel construction $\mathcal{N}_H \otimes \mathcal{A}_e$ two-times, which results in different superactivated asymptotic joint quantum capacities at the channel outputs. Hence, for this channel construction, we obtain radius length $r^*_{(\mathcal{N}_H \otimes \mathcal{N}_H)}$ with weight $p^2$ and we obtain superball radius $r^*_{(\mathcal{N}_H \otimes \mathcal{A}_e)}$ with weight $2p(1-p)$ in $r^*_{super}(\mathcal{N}_1 \otimes \mathcal{N}_2)$. But: using channel combination $\mathcal{N}_H$ and $\mathcal{A}_e$, the term $p^2 r^*_{(\mathcal{N}_H \otimes \mathcal{N}_H)}$ *can never be greater than zero*, because the quantum capacity of the Horodecki channel is zero [52], $Q(\mathcal{N}_H) = r^*_{(\mathcal{N}_H)} = 0$, i.e., radius $r^*_{(\mathcal{N}_H \otimes \mathcal{N}_H)}$ will always have zero length. Now, we focus on superball radius $r^*_{(\mathcal{N}_H \otimes \mathcal{A}_e)}$. The radius $r^*_{(\mathcal{N}_H \otimes \mathcal{A}_e)}$ can be expressed as follows

$$r^*_{(\mathcal{N}_H \otimes \mathcal{A}_e)} \geq \frac{1}{2}\left(\left|\mathbf{r}_1^*\right| + \left|\mathbf{r}_2^*\right|\right), \tag{41}$$

where the radii $\mathbf{r}_1^*$ and $\mathbf{r}_2^*$ of the smallest enclosing quantum informational balls measure the single-use private classical capacities of the channels. As follows, in (41), the $\left|\mathbf{r}_1^*\right|$ and $\left|\mathbf{r}_2^*\right|$ represent the *private classical* capacity of the channels $\mathcal{N}_H$ and $\mathcal{A}_e$, where $P(\mathcal{N}_H) > 0$ and $P(\mathcal{A}_e) = 0$, instead of the quantum capacities $Q(\mathcal{N}_H)$ and $Q(\mathcal{A}_e)$ of the individual channels $\mathcal{N}_H$ and $\mathcal{A}_e$. The radius $r^*_{(\mathcal{N}_H \otimes \mathcal{A}_e)}$ is equal to zero for channel parameters outside the domain $p \notin [0, 0.0041]$.

In Fig. 13, we show the smallest quantum informational balls in the range $0 < p < 0.0041$. In this case, the channels have positive *superactivated* quantum capacity, i.e.,

$$0 < r^*_{(\mathcal{N}_H \otimes \mathcal{A}_e)} \geq \frac{1}{2}\left(\left|\mathbf{r}_1^*\right| + \left|\mathbf{r}_2^*\right|\right) = \frac{1}{2}P^{(1)}(\mathcal{N}_H) = \frac{1}{2}r^{private}_{\mathcal{N}_H}. \tag{42}$$

where $r_{\mathcal{N}_H}^{private}$ is the single-use private classical capacity of the Horodecki channel. The channels $\mathcal{N}_H$ and $\mathcal{A}_e$ with zero quantum capacities individually can be superactivated and a positive capacity can be realized on the output of the channels.

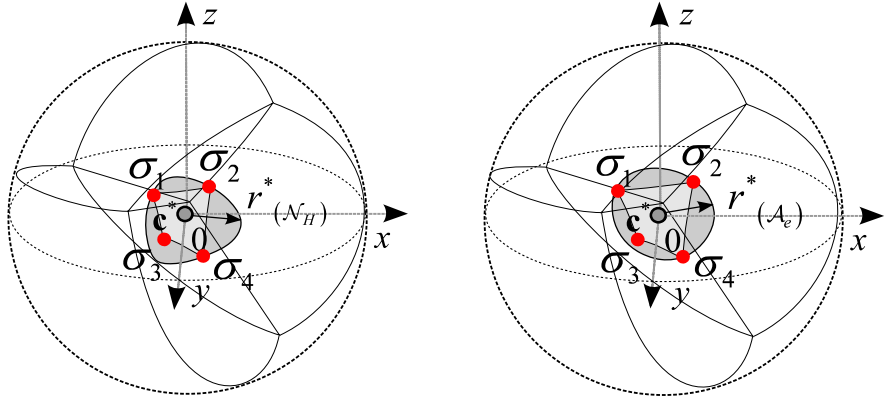

**Fig. 13.** The smallest enclosing balls and radii for Horodecki channel (a) and for erasure channel (b) in the single channel view. If the zero-capacity channels are superactive, the joint capacity will be positive in the given channel parameter domain $0 < p < 0.0041$. (The radii represent the superactivated quantum capacity of the joint structure, using single channel view representation.)

In Fig. 14, we show the smallest enclosing quantum informational balls and the radii of two zero-capacity channels, $\mathcal{N}_1$ and $\mathcal{N}_2$ for channel probabilities $p = 0$ and $p \geq 0.0041$. The radii $r^*_{(\mathcal{N}_H)}$ and $r^*_{(\mathcal{A}_e)}$ are equal to zero for channel parameters outside the domain $0 < p < 0.0041$. The radii $r^*_{(\mathcal{N}_H)}$ and $r^*_{(\mathcal{A}_e)}$ express the *quantum capacities* of the individual channels $\mathcal{N}_H$ and $\mathcal{A}_e$, $Q(\mathcal{N}_H) = Q(\mathcal{A}_e) = 0$, i.e.:

$$r^*_{super}(\mathcal{N}_1 \otimes \mathcal{N}_2) = r^*_{(\mathcal{N}_H \otimes \mathcal{A}_e)} = r^*_{\mathcal{N}_H} + r^*_{\mathcal{A}_e} = 0 \neq \frac{1}{2}(|\mathbf{r}_1^*| + |\mathbf{r}_2^*|). \tag{43}$$

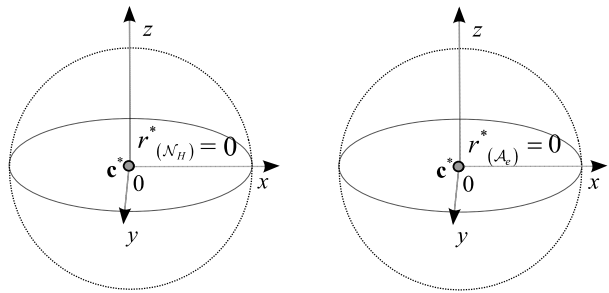

**Fig. 14.** Outputs of two zero-capacity quantum channels (in single channel view). The radii of the smallest quantum informational balls are equal to zero. (The radii represent the quantum capacity of the joint structure, using single channel view representation.)

The results of the superactivation as a function of different $p$ probabilities, where $r^*_{(\mathcal{N}_H \otimes \mathcal{A}_e)}$ is the radius of the superball, which describes the joint capacity of the joint structure $(\mathcal{N}_1 \otimes \mathcal{N}_2)$ are shown in Fig. 15.

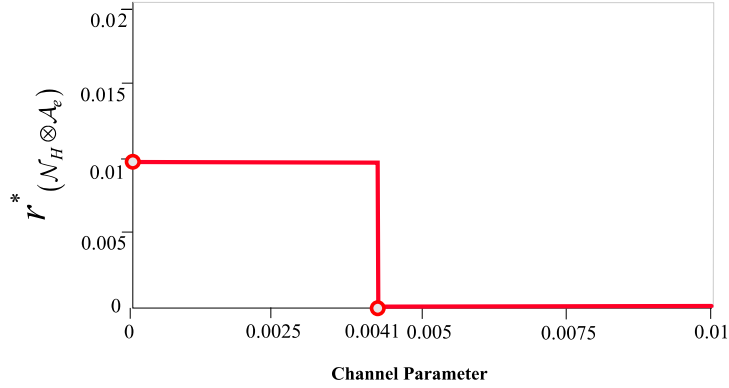

**Fig. 15.** The output of the optimization algorithm describes the radius of the superball, which will be positive only for a given domain of the channel parameter.

As can be observed, the length of the radius of the quantum informational superball is $r^*_{(\mathcal{N}_H \otimes \mathcal{A}_e)} = 0.01$ for channel parameters in the domain $0 < p < 0.0041$.

In Fig. 16, we show we show the length of superball radius $r^*_{super}(\mathcal{N}_1 \otimes \mathcal{N}_2)$, which describes $Q(\mathcal{N}_1 \otimes \mathcal{N}_2)$, see (38). As we have convex combinations of channels, the superball radius $r^*_{super}(\mathcal{N}_1 \otimes \mathcal{N}_2)$ will be measured as $r^*_{(\mathcal{N}_H \otimes \mathcal{N}_H)} = 0$ with zero weight, and $r^*_{(\mathcal{N}_H \otimes \mathcal{A}_e)} = 0.01$ with weight $2p(1-p)$, which lead to $r^*_{super}(\mathcal{N}_1 \otimes \mathcal{N}_2) = 2p(1-p) r^*_{(\mathcal{N}_H \otimes \mathcal{A}_e)} = 2p(1-p) \cdot (0.01)$. We have compared the weights of the radii as a function of channel parameter.

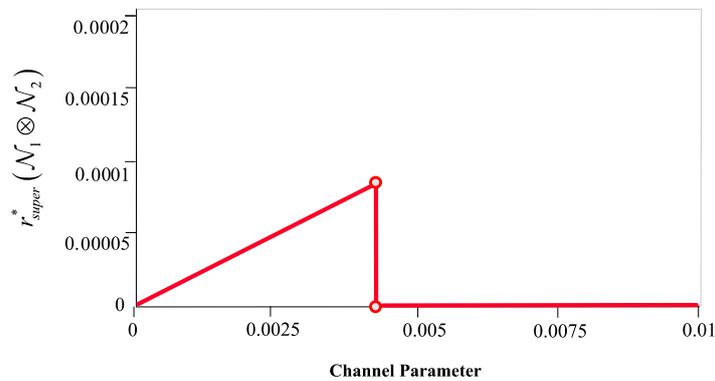

**Fig. 16.** The length of superball radius $r^*_{super}(\mathcal{N}_1 \otimes \mathcal{N}_2)$ as function of the channel parameter.

It can be concluded from our results for a channel parameter in the domain $0 < p < 0.1$ that the output of the algorithm will result in a channel capacity $r^*_{(\mathcal{N}_H \otimes \mathcal{A}_e)} = 0.01$. We have used the sub-domain $0 < p < 0.1$ of parameter $p$ since the critical value is $0 < p < 0.0041$, as found by our algorithm.

The length of the first superball radius is $r^*_{(\mathcal{N}_H \otimes \mathcal{N}_H)} = 0$; however, this output has zero weight. The second superball radius $r^*_{(\mathcal{N}_H \otimes \mathcal{A}_e)}$ has a length of 0.01; this output has a much higher weight, 0.0081, for fixed channel parameter $p = 0.004$, which is the upper bound of the possible range $0 < p < 0.0041$. The maximum weight of the $r^*_{(\mathcal{N}_H \otimes \mathcal{N}_H)}$ can be obtained for channel probability $p = 0.004$.

We can conclude from our numerical analysis that, if we have two zero-capacity channels $\mathcal{N}_1 \otimes \mathcal{N}_2$, then the convex combination of these channels can result in greater than zero capacity for a small subset of possible parameters $p$. As our geometrical analysis revealed, if we have two fixed channel models and we iterate on possible values of parameter $p$, then, from the radius of the smallest superball, we can determine the possible values of the "superactivation parameter". We posit that stronger combinations for superactivation can be constructed from the larger set of quantum channel models and possible parameters.

## 7. Conclusions

This paper exhibited a fundamentally new algorithmic solution for the superactivation of the asymptotic quantum capacity of zero-capacity quantum channels. Using our method, a larger set of "superactive" zero-capacity channels can be discovered very efficiently, and our method can bridge the gap between theoretical and experimental results. To analyze channel superactivation, we introduced the "*superball*" representation, where the radius of the smallest quantum superball is equal to the sum of the radii of the smallest quantum balls of the analyzed quantum channels. The iterations are based on the computed radius of the superball; the iterations are executed on channel input states, channel models, and error probabilities. We have shown that the proposed method is an efficient experimental algorithmic realization of

Smith and Yard's theoretical results, and it can be extended to other channel models and other possible domains of channel probabilities.

In future work, we would like to extend our results to the superactivation of the asymptotic classical zero-error capacity, and we would like to show that other channel combinations can also be superactivated. The proposed algorithmic solution can be the key to finding other possible channel models and channel parameter domains, with possible combinations being proved by theory. If there are other combinations of channel models and channel probabilities that realize superactivation, our method can find them.

## Acknowledgement

The results discussed above are supported by the grant TAMOP-4.2.1/B-09/1/KMR-2010-0002, 4.2.2.B-10/1--2010-0009 and COST Action MP1006.